\documentclass[3p,times]{elsarticle}

\usepackage{ecrc}

\usepackage{epstopdf}

\volume{00}

\firstpage{1}

\journalname{Nuclear Physics A}

\runauth{G.Vujanovic et al.}

\jid{nupha}

\jnltitlelogo{Nuclear Physics A}

\usepackage{graphicx}
\usepackage{amsmath,amssymb}
\usepackage[usenames,dvipsnames,svgnames,table]{xcolor}
\usepackage{soul}

\begin{document}
\setstcolor{red}

\begin{frontmatter}



\title{Probing the non-equilibrium dynamics of hot and dense QCD with dileptons}

\author[McGill]{Gojko Vujanovic}
\author[McGill]{Jean-Fran\c cois Paquet} 
\author[McGill]{Gabriel S. Denicol}
\author[McGill,LBNL]{Matthew Luzum}
\author[BNL]{Bj\"orn Schenke}
\author[McGill]{Sangyong Jeon}
\author[McGill]{Charles Gale}

\address[McGill]{Department of Physics, McGill University, 3600 rue University, Montr\'eal, Qu\'ebec H3A 2T8, Canada}
\address[LBNL]{Lawrence Berkeley National Laboratory, Berkeley, CA 94720, USA}
\address[BNL]{Physics Department, Brookhaven National Lab, Building 510A, Upton, NY, 11973, USA }

\begin{abstract}
It is argued that, in heavy ion collisions, thermal dileptons are good probes of the transport properties of the medium created in such events, and also of its early-time dynamics, usually inaccessible to hadronic observables. In this work we show that electromagnetic azimuthal momentum anisotropies do not only display a sensitivity to the shear relaxation time and to the initial shear-stress tensor profile, but also to the temperature dependence of the shear viscosity coefficient.
\end{abstract}

%\begin{keyword}
%Keyword1 \sep Keyword2 \sep Keyword3
%\end{keyword}

\end{frontmatter}

% \linenumbers


%%%%%%%%%%%%%%%%%%%%%%%%%%%%%%%%%%%%%%%%%%%%%%%%%%%%%%%%%%%%%%%%%%%%%

\section{Introduction}
 
%%%%%%%%%%%%%%%%%%%%%%%%%%%%%%%%%%%%%%%%%%%%%%%%%%%%%%%%%%%%%%%%%%%%%
One of the main goals of Relativistic Heavy Ion Colliders, either  the Relativistic Heavy Ion Collider (RHIC, at Brookhaven National Laboratory) or the Large Hadron Collider (LHC, at CERN), is to investigate the thermodynamic and transport properties of the hot and dense phase of QCD. Much work has been concentrated on the determination of an effective value of the shear viscosity coefficient from analyses of relativistic heavy-ion collisions but so far, such investigations haven been performed mostly by comparing to hadrons produced at the final stages of the collision. 

Electromagnetic radiation constitutes a class of complementary and penetrating probes that are sensitive to the entire space-time history of nuclear collisions including its very early stages. In this contribution we show that thermal dileptons are affected by the transport properties of the fluid and by the non-equilibrium aspects of the initial state that are usually inaccessible to hadronic probes. We establish that the azimuthal momentum anisotropies of thermal dileptons are particularly sensitive to the temperature dependence of the shear viscosity coefficient. We also show the potential of thermal dileptons in differentiating between possible initial shear-stress tensors and shear relaxation times.
%%%%%%%%%%%%%%%%%%%%%%%%%%%%%%%%%%%%%%%%%%%%%%%%%%%%%%%%%%%%%%%%%%%%%

\section{Fluid-dynamical model }

%%%%%%%%%%%%%%%%%%%%%%%%%%%%%%%%%%%%%%%%%%%%%%%%%%%%%%%%%%%%%%%%%%%%%
We will discuss only Au-Au collisions at $\sqrt{s_{NN}}=200$ GeV. The time evolution of the hot and dense medium created at RHIC is modeled using \textsc{music}, a 3+1D hydrodynamical evolution \cite{Schenke:2010rr}. The main equations of motion are the conservation laws of energy and momentum, given by the continuity equation for the energy momentum tensor, $T^{\mu \nu }$, i.e., $\partial _{\mu}T^{\mu \nu }=0$. As usual, $T^{\mu \nu }=\varepsilon \,u^{\mu }u^{\nu}-\Delta ^{\mu \nu }P+\pi ^{\mu \nu }$, with $\varepsilon $ being the energy density, $P$ the thermodynamic pressure, $u^{\mu }$ the fluid four-velocity, $\pi ^{\mu \nu }$ the shear-stress tensor, and $\Delta ^{\mu \nu }=g^{\mu\nu }-u^{\mu }u^{\nu }$ the projection operator onto the 3-space orthogonal to the velocity, with a metric tensor  $g^{\mu \nu} = {\rm diag}(1, -1, -1, -1).$ The lattice QCD equation of state is used to relate $P$ and $\varepsilon$ \cite{Huovinen:2009yb}.

The conservation laws are complemented by a relaxation equation for the shear-stress tensor, given by a version of Israel-Stewart (I-S) theory \cite{Israel1976310,Israel:1979wp}, 
\begin{equation}
\tau _{\pi }\Delta _{\alpha \beta }^{\mu \nu }u^{\lambda }\partial _{\lambda}\pi ^{\alpha \beta }+\frac{4}{3}\tau _{\pi }\pi ^{\mu \nu }\partial_{\lambda }u^{\lambda }=\left( \pi _{\mathrm{NS}}^{\mu \nu }-\pi ^{\mu \nu}\right) \text{,}  \label{eq:pi_munu}
\end{equation}%
where $\pi _{\mathrm{NS}}^{\mu \nu }=2\eta \,\sigma ^{\mu \nu }=2\eta \Delta_{\alpha \beta }^{\mu \nu }\partial ^{\alpha }u^{\beta }$ is the Navier-Stokes limit of the shear-stress tensor, with $\Delta _{\alpha \beta}^{\mu \nu }=\left( \Delta _{\alpha }^{\mu }\Delta _{\beta }^{\nu }+\Delta_{\beta }^{\mu }\Delta _{\alpha }^{\nu }\right) /2-\Delta _{\alpha \beta}\Delta ^{\mu \nu }/3$ being the double, symmetric, traceless projection operator. In its simplest from, I-S theory has two transport coefficients: the shear viscosity $\eta $, also present in Navier-Stokes theory, and the shear relaxation time, $\tau _{\pi }$, which only exists in I-S theory.

We use a constant value $\eta/s=1/4\pi$ as the default value for the shear viscosity over entropy density ratio. In the QGP phase, i.e. for temperatures  above a transition temperature $T_{\mathrm{tr}}=0.18$ GeV, we will also consider an $\eta /s$ with linear temperature dependences of the form $\eta /s(T)=a(T/T_{\mathrm{tr}}-1)+1/4\pi $. The effect of the temperature dependence of $\eta /s$ on hadronic and eletromagnetic flow observables is tested by modifying the slope parameter $a$. The values of $a$ employed in this work are $a=0$, $0.2427$, and $0.5516$, with $a=0$ corresponding to the constant default value. 

The shear relaxation time is assumed to be of the form $\tau _{\pi }=b_{\pi}\eta /\left( \varepsilon +P\right) $. The role of $\tau _{\pi }$ is to govern the rate at which $\pi ^{\mu \nu }$ evolves and relaxes towards its Navier-Stokes limit. The default value used in this study is $b_{\pi }=5$. Here, we test the effect of larger relaxation times by also considering $b_{\pi }=$ $10$ and $20$.

The initial energy density profile is determined by the Monte-Carlo Glauber model, with all the free parameters being tuned to describe the multiplicity and elliptic flow of hadronic observables at RHIC's highest energy. The initial value of the shear-stress tensor is also varied in this work and is parametrized in the following way $\pi _{0}^{\mu \nu }=c\times2\eta \sigma ^{\mu \nu }$. The parameter $c$ controls the deviation of the initial state from local thermodynamic equilibrium. Here, we set $c=0$, 1/2, and 1, with the default value being zero. The initial velocity profile is always set to zero in hyperbolic coordinates.

%%%%%%%%%%%%%%%%%%%%%%%%%%%%%%%%%%%%%%%%%%%%%%%%%%%%%%%%%%%%%%%%%%%%%

\section{Thermal dilepton rates}

%%%%%%%%%%%%%%%%%%%%%%%%%%%%%%%%%%%%%%%%%%%%%%%%%%%%%%%%%%%%%%%%%%%%%
Thermal dilepton rates can generically be expressed as: 
\begin{eqnarray}
\frac{d^4 R}{d^4 q} = -\frac{\alpha}{12\pi^4} \frac{1}{M^2} \frac{{\rm Im} \Pi^{{\rm R}}_{\gamma^{\ast}}}{e^{ q^0/T}-1}
\label{eq:rate}
\end{eqnarray}
where $\alpha $ is the electromagnetic structure constant, $\Pi _{\gamma^{\ast}}^{\mathrm{R}} = \Pi^{\mathrm{R},\,\, \mu}_{\gamma^{\ast},\,\mu}$ is the trace of the retarded virtual photon self-energy, and $M^{2}=q^{2}$, where $M$ is the virtual photon invariant mass. This expression is valid at leading order in $\alpha_{\rm em}$, but is exact at all orders of $\alpha_{\rm s}$ \cite{kapusta-gale-book}. We have used the Born rate in this paper, which corresponds to the quark-antiquark annihilation rate into dileptons. Viscosity is included via a deviation of the thermal distribution functions $n$ entering in evaluating $\Pi _{\gamma ^{\ast }}^{\mathrm{R}}$ such that $n\rightarrow n+\delta n$, where $\delta n(p)=G(p)n(p)(1\pm n(p))p_{\mu}p_{\nu }\pi ^{\mu \nu }/\left[ 2T^{2}\left( \varepsilon +P\right) \right]$, and $G(p)$ is a function that must be determined through the use of microscopic physics. In order to find $G(p)$, we solved the Boltzmann equation assuming a massless gas of particles with constant $2\rightarrow 2$ cross section. 

The general form of the thermal dilepton rates Eq. (\ref{eq:rate}) can be applied in the hadronic sector (low temperatures) via the introduction of the Vector Dominance Model (VDM). Through VDM, $\Pi _{\gamma^{\ast }}^{\mathrm{R}}$ is expressed in terms of $D_{V}^{\mathrm{R}}$, the vector meson $(V)$ retarded propagator. A key ingredient in evaluating $D_{V}^{\mathrm{R}}$ is the vector meson self-energy $\Pi _{V}$, the latter being presented in detail in Ref.~\cite{Vujanovic:2013jpa}.

%%%%%%%%%%%%%%%%%%%%%%%%%%%%%%%%%%%%%%%%%%%%%%%%%%%%%%%%%%%%%%%%%%%%%

\section{Results}

%%%%%%%%%%%%%%%%%%%%%%%%%%%%%%%%%%%%%%%%%%%%%%%%%%%%%%%%%%%%%%%%%%%%%
\begin{figure}[!h]
\begin{center}
\begin{tabular} {c c}
\includegraphics[width=0.45\textwidth]{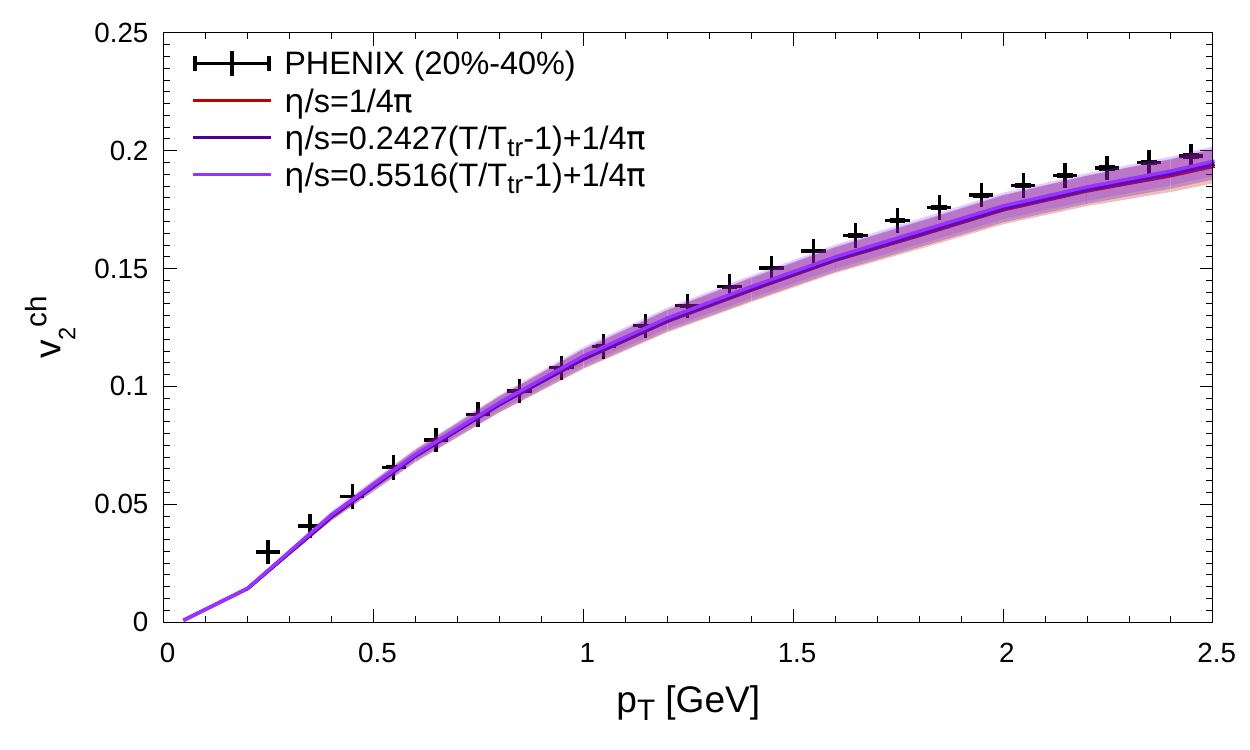} \includegraphics[width=0.45\textwidth]{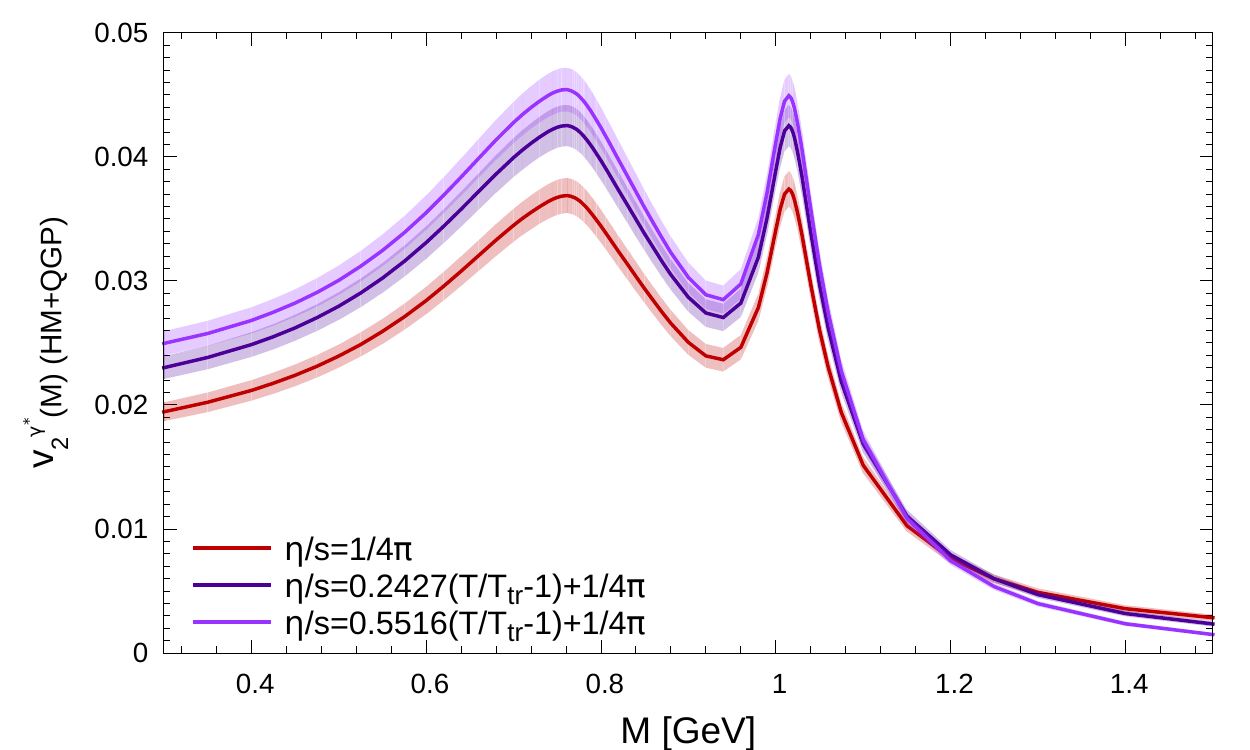}
\end{tabular}
\end{center}
\caption{The effects of varying $\frac{\eta}{s}(T)$ in the QGP phase on the elliptic flow of charged hadrons (left panel) and of virtual photons (right panel),  in collisions of Au + Au at 200 $A$ GeV, at a 20--40\% centrality class.}
\label{fig:v2_eta_s_T}
\end{figure}

\begin{figure}[!h]
\begin{center}
\begin{tabular} {c c}
\includegraphics[width=0.45\textwidth]{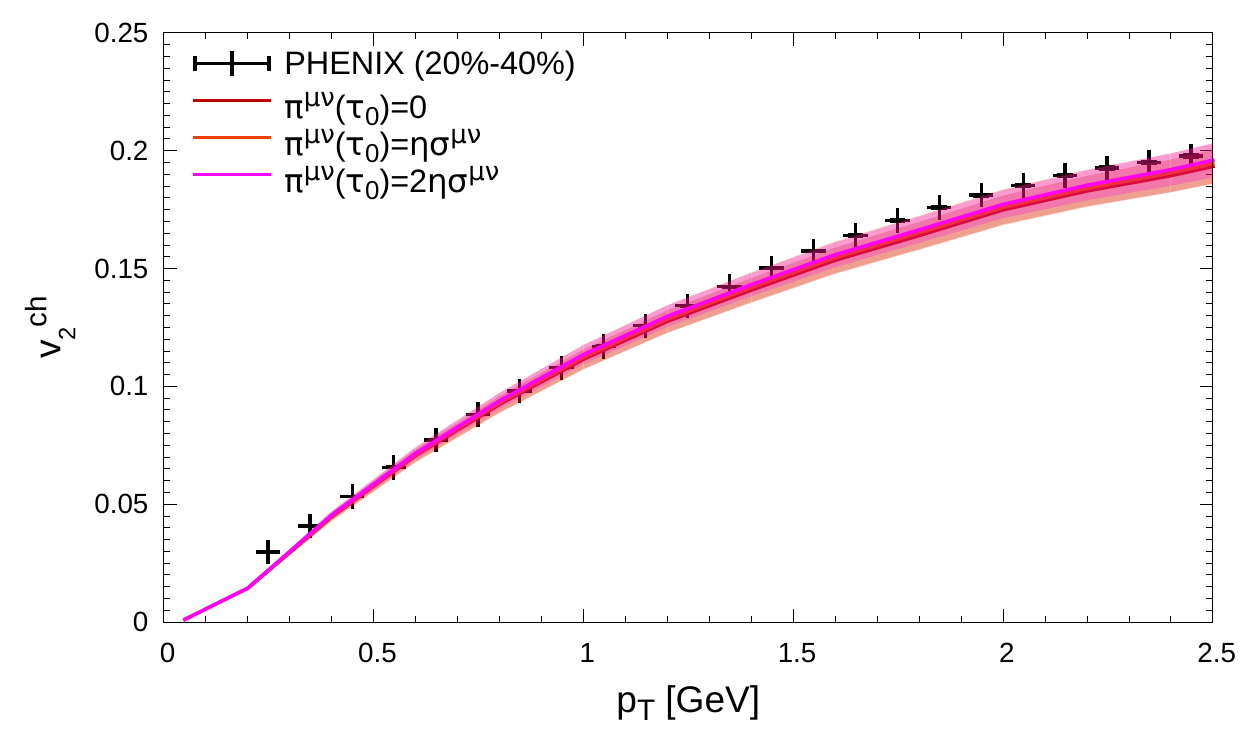} \includegraphics[width=0.45\textwidth]{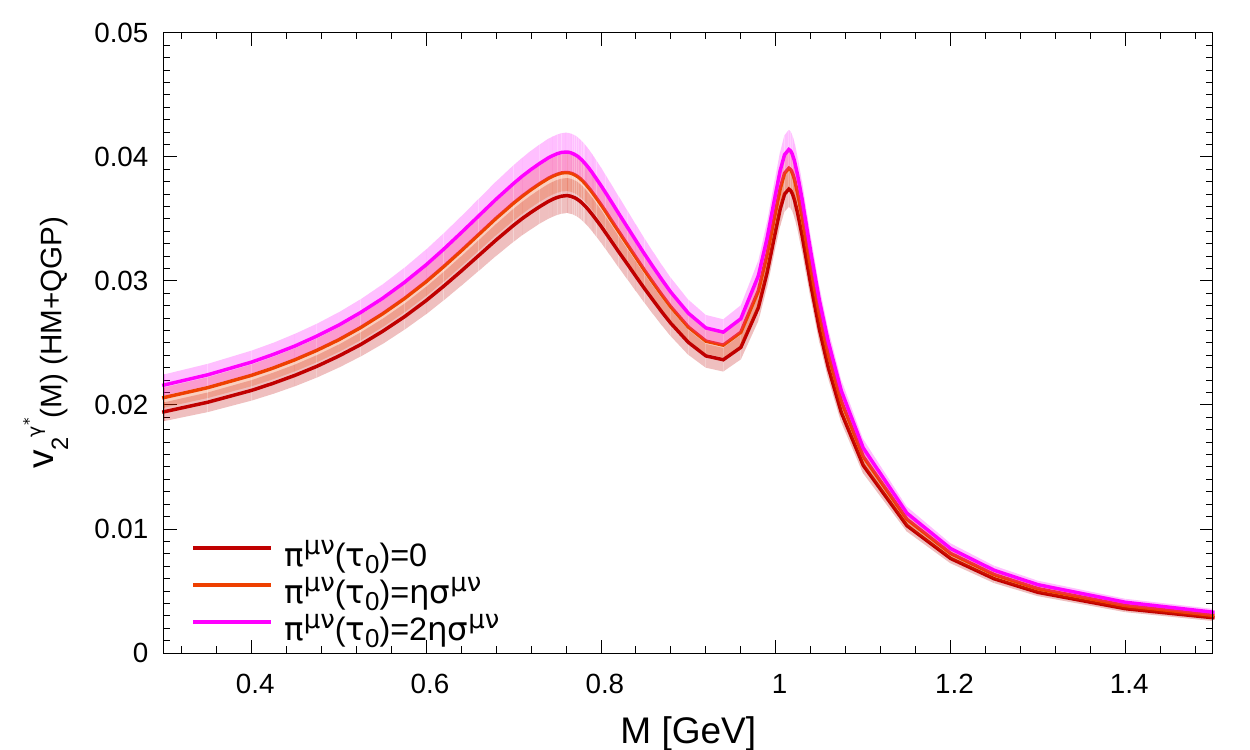}
\end{tabular}
\end{center}
\caption{The effects of varying $\pi^{\mu\nu}_0$ on charged hadron's (left panel) and virtual photon's (right panel) elliptic flow created in collisions of Au + Au at 200 $A$ GeV, in the  20--40\% centrality class.}
\label{fig:v2_init_cond}
\end{figure}

\begin{figure}[h]
\begin{center}
\begin{tabular}{cc}
\includegraphics[width=0.45\textwidth]{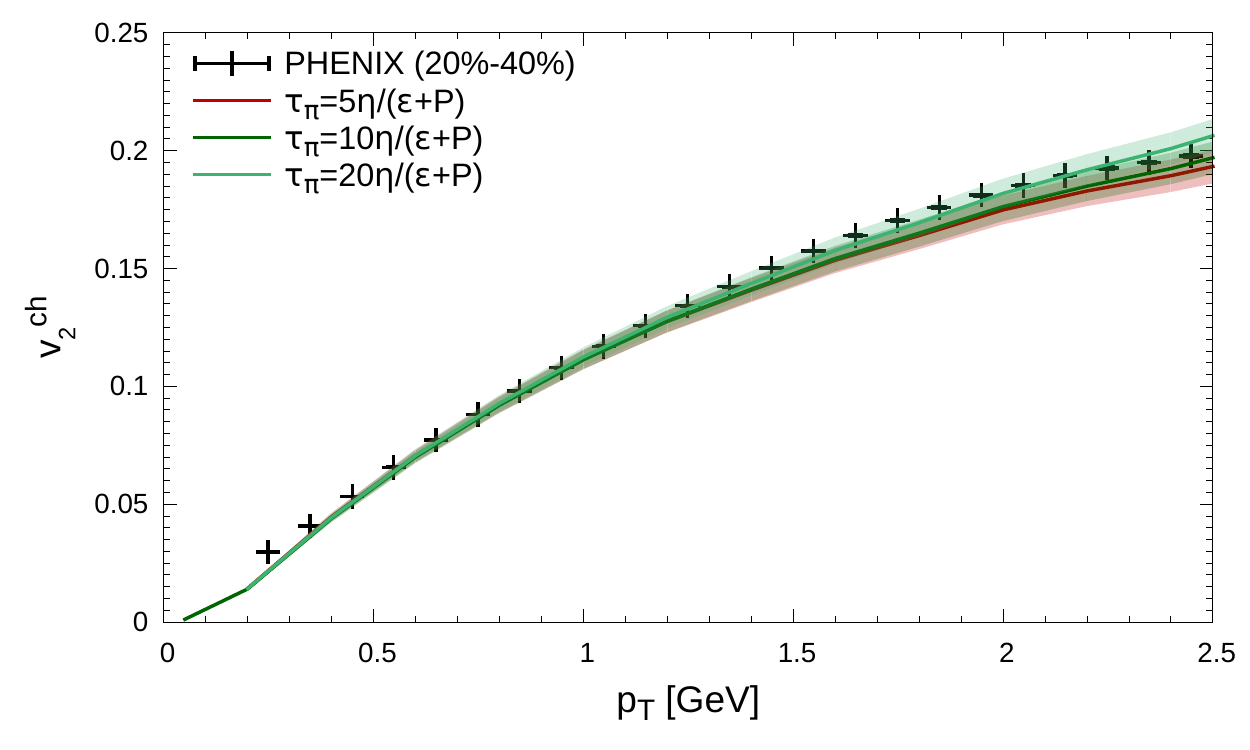}\includegraphics[width=0.45\textwidth]{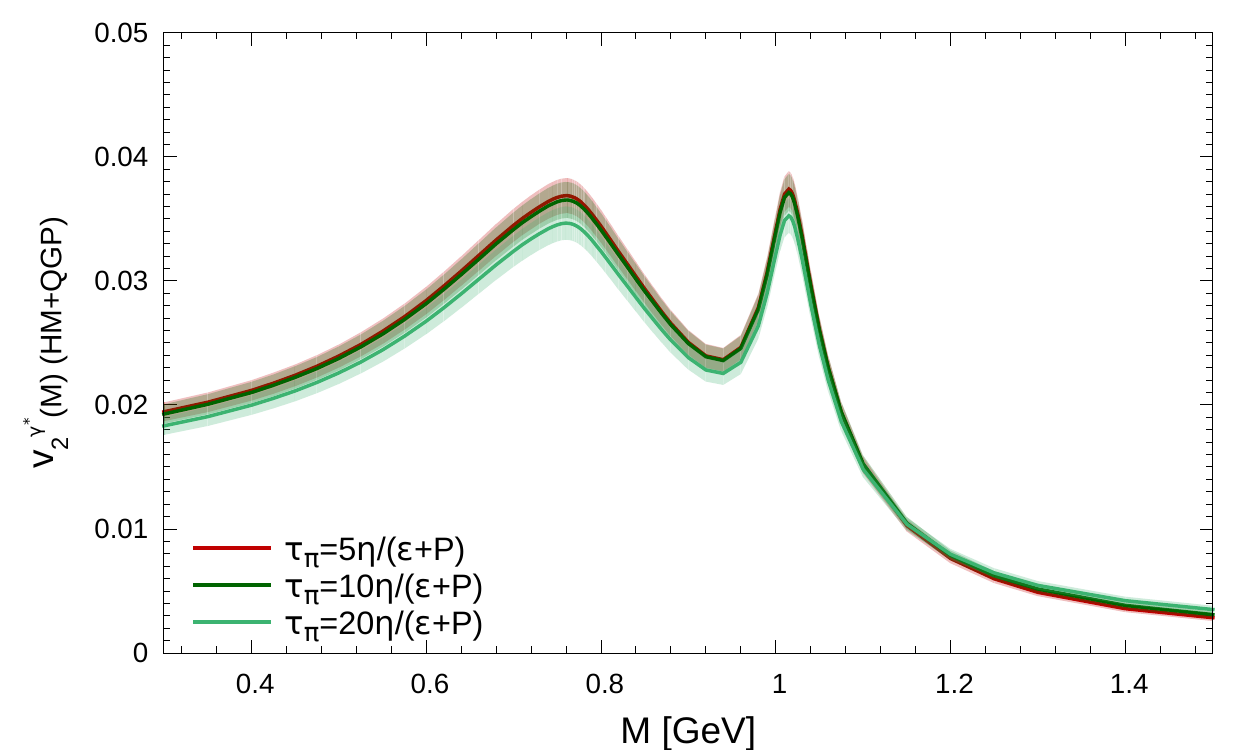} & 
\end{tabular}
\end{center}
\caption{The effects of varying $\protect\tau _{\protect\pi }$ on charged
hadron's (left panel) and virtual photon's (right panel) elliptic flow
created in collisions of Au + Au at 200 $A$ GeV, in the 20--40\% centrality
class.}
\label{fig:v2_tau_pi}
\end{figure}

In the left panels of Figures~\ref{fig:v2_eta_s_T}, \ref{fig:v2_init_cond}, and \ref{fig:v2_tau_pi}, we show the differential elliptic flow of charged hadrons as a function of transverse momentum, $v_{2}(p_{T})$. In the right panel of the same figures we show the integrated elliptic flow of thermal dileptons as a function of their invariant mass, $v_{2}(M)$. In Figure~\ref{fig:v2_eta_s_T} $\eta/s$ was varied, while in Figures~\ref{fig:v2_init_cond} and \ref{fig:v2_tau_pi} $\pi _{0}^{\mu \nu }$ and $\tau _{\pi }$ were changed, respectively. In each case, the parameters that are not varied are kept at their default values. For each parameter configuration, we computed 200 events, all in the 20--40\% centrality class. The color bands in the plots indicate the statistical uncertainties of the calculations. We note that  our results for charged hadron $v_2(p_{T})$ are in good agreement with PHENIX data, which corresponds to the points in the left panels of our figures.

It was already shown in Ref.~\cite{Niemi:2011ix} that charged hadrons have a small sensitivity to the $\eta /s(T)$ in the QGP phase at the top RHIC energy. The left panel of Figure \ref{fig:v2_eta_s_T} illustrates that this behavior still holds true for the temperature dependence of $\eta /s$ used in this study. In addition, the results plotted in the left panels of Figures \ref{fig:v2_init_cond}  and \ref{fig:v2_tau_pi} confirm that the elliptic flow of charged hadrons at RHIC's highest energy has a very small sensitivity also to variations of $\pi _{0}^{\mu \nu }$ and of $\tau _{\pi }$. Even though it is not shown here, we verified that the same is true for the transverse momentum spectra of charged hadrons.

The situation is not the same when it comes to thermal dileptons. The effect of  varying $\pi _{0}^{\mu \nu }$ and $\tau _{\pi }$ is visible on $v_{2}(M)$, but it is still relatively modest, as seen in the right panels of Figures \ref{fig:v2_init_cond} and \ref{fig:v2_tau_pi}. However, the magnitude of the slope of $\eta /s$ as a function of $T$ has a sizeable influence on the elliptic flow of thermal dileptons; this is shown in the right panel of Figure \ref{fig:v2_eta_s_T}. Recall that, unlike the charged hadrons emitted at the freeze-out hyper-surface, thermal dileptons are emitted throughout the collision history and their elliptic flow retains a memory of the $\eta /s(T)$ in the QGP phase. For small and intermediate values of invariant mass ($M<1.2$ GeV), increasing the QGP's $\eta /s(T)$ leads to an increase in $v_{2}(M)$. Meanwhile, for larger values of invariant mass ($M>1.2$ GeV) the behavior is inverted and a larger $\eta /s(T)$ leads to a smaller elliptic flow coefficient. We have found that this inversion also occurs for the momentum anisotropy, $\epsilon_p=\langle T^{xx}-T^{yy}\rangle/\langle T^{xx}+T^{yy}\rangle$, when plotted against time.        

The change in the hydrodynamical evolution induced by a T-dependent $\eta /s$ is occuring far from the freeze-out surface and therefore is only accessible to electromagnetic probes. At freeze-out (for collisions at RHIC energies), most of the memory of different values of $\eta/s$ in the QGP phase has faded: the charged hadrons $v_2$ is thus mostly unaffected (see left panel of Figure \ref{fig:v2_eta_s_T} and Ref. \cite{Niemi:2011ix}).

%%%%%%%%%%%%%%%%%%%%%%%%%%%%%%%%%%%%%%%%%%%%%%%%%%%%%%%%%%%%%%%%%%%%%

\section{Conclusions}

%%%%%%%%%%%%%%%%%%%%%%%%%%%%%%%%%%%%%%%%%%%%%%%%%%%%%%%%%%%%%%%%%%%%%

 In this contribution, we showed that thermal dileptons are affected by the transport properties of the QGP and by non-equilibrium aspects of the initial evolution that are usually inaccessible to hadronic probes. For the first time, we explicitly demonstrate that the invariant mass distribution of dileptons and their azimuthal momentum anisotropy have a small but non-negligible dependence on the magnitude of the shear relaxation time and on the value of initial shear-stress tensor. Importantly, virtual photons may also reveal the temperature dependence of the shear viscosity coefficient. This endeavor reaffirms the potential that penetrating probes, such as dileptons, have in furthering  our understanding of QCD at high temperatures and densities. We expect that, as experimental uncertainties become smaller, such probes will play a more dominant role in the extraction of the initial state and transport properties of the bulk QCD matter created in ultrarelativistic heavy ion collisions at RHIC and at the  LHC.

\section{Acknowledgements}
This work was supported in part by the Natural Sciences and Engineering Research Council of Canada, in part by U. S. DOE Contract No. DE-AC02-98CH10886, and in part by the by the Director, Office of Energy Research, Office of High Energy and Nuclear Physics, Division of Nuclear Physics, of the U.S. Department of Energy under Contract No. DE-AC02-05CH11231. G. Vujanovic acknowledges support by the Canadian Institute for Nuclear Physics, and G.S. Denicol acknowledges support through a Banting Fellowship from the Government of Canada. Computations were performed on the Guillimin supercomputer at McGill University under the auspices of Calcul Qu\'ebec and Compute Canada. The operation of Guillimin is funded by the Canada Foundation for Innovation (CFI), the National Science and Engineering Research Council (NSERC), NanoQu\'ebec, and the Fonds Qu\'eb\'ecois de Recherche sur la Nature et les Technologies (FQRNT).



%% References


\bibliographystyle{elsarticle-num}
\bibliography{references}

\begin{thebibliography}{1}
\expandafter\ifx\csname url\endcsname\relax
  \def\url#1{\texttt{#1}}\fi
\expandafter\ifx\csname urlprefix\endcsname\relax\def\urlprefix{URL }\fi
\expandafter\ifx\csname href\endcsname\relax
  \def\href#1#2{#2} \def\path#1{#1}\fi

\bibitem{Schenke:2010rr}
B.~Schenke, S.~Jeon, C.~Gale, Phys.Rev.Lett. 106 (2011) 042301.

\bibitem{Huovinen:2009yb}
P.~Huovinen, P.~Petreczky, Nucl.Phys. A837 (2010) 26--53.

\bibitem{Israel1976310}
W.~Israel, Annals of Physics 100~(1–2) (1976) 310 -- 331.

\bibitem{Israel:1979wp}
W.~Israel, J.~Stewart, Annals Phys. 118 (1979) 341--372.

\bibitem{kapusta-gale-book}
J.~I. Kapusta, C.~Gale, {\it Finite-temperature field theory: Principles and
  applications}, Cambridge, UK, University Press.

\bibitem{Vujanovic:2013jpa}
G.~Vujanovic, C.~Young, B.~Schenke, R.~Rapp, S.~Jeon, et~al., Phys. Rev. C 89
  (2014) 034904.

\bibitem{Niemi:2011ix}
H.~Niemi, G.~S. Denicol, P.~Huovinen, E.~Molnar, D.~H. Rischke, Phys.Rev.Lett.
  106 (2011) 212302.

\end{thebibliography}

\end{document}